\documentclass[useAMS,usenatbib]{mn2e}
\usepackage{graphicx}
\usepackage{journals}

\title[Chromospheric plasma ejection above a pore]{Chromospheric plasma ejection above a pore}
\author[L. Bharti et al.]
{L. Bharti$^{1}$ \thanks{E-mail:lokesh$\_$bharti@yahoo.co.in},
B. Shobha$^{2}$, C. Quintero Noda$^{3,4}$, C. Joshi$^{2}$, U. Pandya$^{5}$ \\
$^{1}$Bal Shiksha Sadan Society, 3 Ratakhet, Sajjan Nagar, Udaipur, Rajasthan, India\\
$^{2}$JECRC University, Jaipur, Rajasthan, India\\
$^{3}$Rosseland Centre for Solar Physics, University of Oslo, P.O. Box 1029 Blindern, N-0315 Oslo, Norway\\
$^{4}$Institute of Theoretical Astrophysics, University of Oslo, P.O. Box 1029 Blindern, N-0315 Oslo, Norway\\
$^{5}$Pecific University, Udaipur, Rajasthan, India\\
}
\begin{document}

\date{Accepted 1988 December 15. Received 1988 December 14; in original form 1988 October 11}

\pagerange{\pageref{firstpage}--\pageref{lastpage}} \pubyear{2013}

\maketitle

\label{firstpage}

\begin{abstract}
We present high spatial resolution observations of short lived transients, ribbon and jets like events above a pore in Ca II H images where
fine structure like umbral dots, lightbridge and penumbral micro filaments are present in the underlying photosphere. We found that current layers are formed at the edges of convective fine structure due to the shear between their horizontal field and the ambient vertical field. High vertical electric current density patches are observed in the photosphere around these events which indicates the formation of a current sheet at the reconnection site.
In the framework of past studies, low altitude reconnection could be the mechanism that produces such events.
The reconnection is caused by an opposite polarity field produced by the bending
of field lines by convective downflows at the edge of the pore fine structures.
\end{abstract}


\section{Introduction}

Magnetic flux concentrations on the solar surface are observed as a wide range of features. In terms of diameter, we have sub-arc sec magnetic elements like micropores that grow in size as pores (few Mm) up to sunspots (more than 40 Mm).
The main feature of a pore is the absence of penumbra which is interpreted
in terms of vertical magnetic field in a magnetostatic flux tube \cite[see][and references therein]{sobotka12}.
Pores appear on the surface when the field strength of magnetic elements becomes larger than 650 Gauss and the intensity drops lower than 0.85 of the quiet surrounding
photosphere \citep{dorotovi16}. From a larger sample of pores, we have that the mean intensity could be up to 40$\%$ lower than the surrounding quiet-Sun granulation \citep{verma14}. Interestingly, fine structures such as umbral dots (UDs) and lightbridges which are observed in sunspots are also present in some pores. \cite{bharti16} reported the existence of darkcored filaments at the edges of granules facing the core of the pore, similar to penumbral filaments in sunspots. This indicates a similarity
between convective processes happening in pores and sunspots \citep{giordano08,ortiz10}.

Lightbridges (LBs) are bright long structure that separates sunspot umbra of same polarity.
They appear during the formation and the decay phase of a sunspot evolution. Typically, there are two types of
LBs : granular and filamentary \citep{sobotka94}. The magnetic filed in LBs is weaker than their surrounding and more horizontal (Leka 1997). Recent
observational findings suggest that the nature of LBs and UDs is magnetoconvective \citep{bharti07a,bharti09,rimmele08}. Realistic 3D MHD simulations
of active region emergence also suggest common magnetoconvective origin of LBs and UDs \citep{cheung10,toriumi15b}. Plasma ejections or surge like activity above sunspot lightbridges in the chromosphere have been observed \cite[see][and references therein]{shimizu09}.
These events are explained in terms of low altitude reconnection between preexisting umbral filed and newly emerging magnetic field.
\cite{shimizu09} found strong vertical electric current patches of opposite sign at the interface between the current-carrying
flux tube (LB) and the pre-existing vertical umbral field. This suggest the formation of a current sheet at the reconnection site. However, the rising of current-carrying flux tubes in light bridges is not supported either by observations (e.g. Robustini et al. 2016) or by simulations (e.g. Toriumi et al. 2015), although these authors found electric currents, the indications of a current sheet. LBs are also observed in pores and also show enhanced heating in the chromosphere, similar to sunspots LBs \citep{sobotka13}.

The chromosphere above sunspots and pores umbra is very dynamic in terms of umbral flashes, 3-min oscillations (Sobotka et al. 2013 and references therein). \cite{keys18} studied the oscillatory behaviour of pores and found that the
surface mode is more dominant than the body mode and that an oscillatory energy flux could affect the atmospheric dynamics above pores. \cite{bharti13} reported the presence of jet like events in Ca II H observations above a sunspot umbra. Co-spatial
and co-temporal photospheric observations show that some of those jet-like events are above UDs. They interpreted
such events as the result of a magnetic reconnection between opposite polarity fields caused by strong convective downflows.
However, this scenario is seen only in larger UDs that appears in MHD simulations \citep{bharti10b}.
No correlation between occurrence of umbral dots and umbral flashes with spatial location of small-scale umbral brightenings above sunspot umbra were found by \cite{nelson17}. Observational evidence for one to one correspondence of underlying UDs in the photosphere and jets in the upper atmosphere is still elusive.

In this paper we use G-band, Ca II H and spectropolarimetric measurements to present
plasma ejections above small scale dynamic fine structures of a pore umbra and their underlying magnetic field.

\section[]{Observations and data analysis}

The pore studied in detail was observed in the NOAA Active Region 10955 on May 11, 2007. The active region was located close to the disc center at a heliocentric angle of $\mu$=0.99. It shows a larger pore and several smaller pores and micropores (see Figure 1). The observations were taken with the Broadband Filter Imager (BFI) attached to the Solar Optical Telescope (SOT) \citep{tsuneta08} on board the Hinode satellite (Kosugi et al. 2007). BFI recorded images from 18:08:42 UT to 23:38:04 UT in the G-band (4305\,\AA, bandwidth: 8 \,\AA) and Ca II H (3968 \,\AA, bandwidth: 3 \,\AA) line with a cadence of 30 seconds. The image scale is 0\farcs054 per pixel and a spatial resolution of 0\farcs22 and 0\farcs20 is achieved for both wavelengths, respectively. We applied a Wiener filter \citep{sobotka93} to all the images of both wavelengths to correct the point spread function (PSF) of the telescope assuming diffraction limit on an ideal 50 cm circular aperture. All images of both time series were
spatially coaligned using a cross-correlation algorithm. This active region was also observed with the Spectropolarimeter (SP) \citep{splites13} on board Hinode/SOT which recorded the Stokes I, Q, U, and V profiles for the Fe I 6301.5 and 6302.5 \AA~ spectral lines. We used two fast maps before and after the plasma ejection event. The integration time for the fast maps was 1.6 sec.
The spectral sampling is 21.549 m\AA ~pixel$^{-1}$.
The field of view a) comprises a square are of 163\farcs84
b) comprises an area of 151\farcs36 $\times$ 1630\farcs84. The spatial
sampling for the fast map was 0\farcs316 along the slit and 0\farcs295 in
the scanning direction. The spatial resolution of the resulting spectropolarimetric
map is approximately 0\farcs6 for the fast maps.
The calibration of the SP data is described by \cite{ichimoto08}.
We used the Solar-Soft pipeline to calibrate the SP data.

The spatial deconvolution of the observed data was performed using the method presented in \citet{ruizcobo13} and \citet{qnoda15}.
To reconstruct each Stokes profiles (I, Q, U, V) we used (8, 5, 5, 8) eigenvectors of principal component analysis (PCA). The total number of
15 iteration steps were performed in the deconvolution process which increased the continuum contrast to 11.51$\%$ per cent while it was 7.1$\%$ per cent in the original data. At this step the iteration process was stopped because it provides an increase factor of the continuum contrast similar to the one that was obtained in previous works \citep{qnoda15,qnoda16a}.

We inferred the properties of the solar atmosphere using the inversion code SIR (Stokes Inversion based on Response functions; Ruiz Cobo \& del Toro Iniesta (1992)). The code allows us to determine the atmospheric parameters at different optical depths for each pixel independently. We use a single magnetic component in the inversion process, and no stray light contribution is included because we assume the deconvolution process has corrected it. We define the number of nodes of each atmospheric parameter automatically as it is explained in Quintero Noda et al. (2016). We limited the maximum number of nodes to five for the gas temperature, three for the line-of-sight (LOS) velocity, three for the magnetic field strength and inclination, one for the magnetic field azimuth, and also one for the microturbulence. We reproduce the spectral characteristics of the instrument convolving in each iteration the SIR synthetic profile with the Hinode/SP spectral  PSF (Lites et al. 2013). We start the inversion of each pixel using the HSRA atmosphere (Gingerich et al. 1971) as a guess model following the strategy described in Quintero Noda et al. (2015).

\section{Results}

\begin{figure}
 \hspace{-8mm}
 \includegraphics[width=180mm]{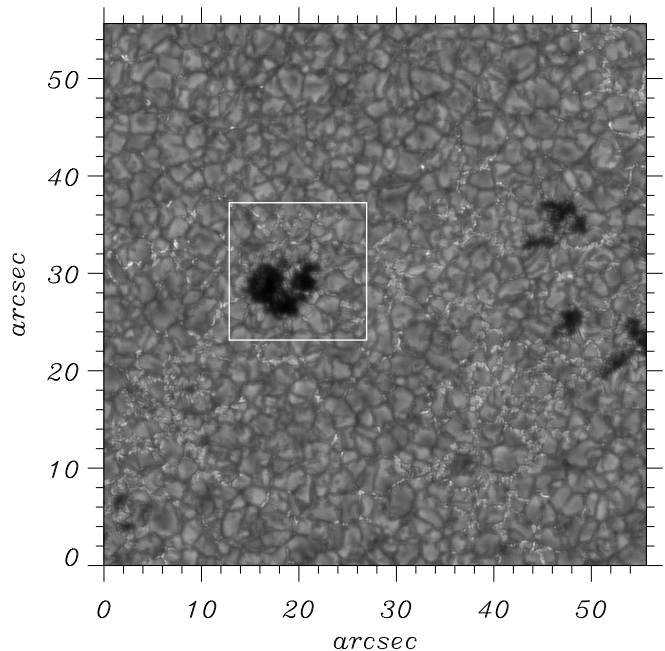}
 \vspace{-10mm}
 \caption{G-band image observed by the Broadband Filter Imager (BFI)
 on Hinode, SOT of the active Region NOAA 10955 on 11 May 2007 at 18:08 UT. The center of the field of view is located at the 09 S 06 W. White rectangle indicate the pore studied in detail.}
\end{figure}

\subsection{Overall evolution in the photosphere}
The pore  of  interest was observed in G-band and Ca II H from 18:08 UT to 23:38 UT. In both time series there are three gaps where the data is not available, i.e. from 18:42-19:08, 20:21-20:47, and 21:59-22:26 UT. Figure 1 illustrates the first image
of the G-band time series. Several pores, micropores as well as brightpoints are also visible. The pore enclosed by the white box displays lightbridge fragments. In the following we discuss its evolution in detail (see Fig. 2).

At 18:08 several lightbridge fragments are visible. The fragments
adjacent to the edges of the pore are dispersed. Later the lightbridge fragment in the center of the pore increases in size. At 19:13 UT
an isolated lightbridge is visible. This lightbridge changes shape and develops in to a larger cell with multiple
dark lanes. During this transition from lightbridge fragments to a larger cell, we notice that the lightbridge
detached from the pore boundaries. This development is illustrated in subsequent images till 21:06 UT in Figure 2. From 21:44 UT to the end of the time series the larger cell transforms into a thin lightbridge again dividing the magnetic structure in two parts.  A better view of the evolution can be found in the MPEG movie available as online material. During the time series the pore sustains its identity with the main changes related to the lightbridge and cell features.

A close inspection of the movie shows that the larger cell displays plasma motions from its center towards the edges along the darklanes. This is a typical characteristic of convection where the hot material raises in the center of the cell and sinks at the edges. Thus, this larger cell has similar properties as that found in larger UDs. This is in agreement with \cite{bharti09} where the authors showed that such larger cells display all the properties of normal UDs. However, in comparison to traditional umbral dots, the present structure has a larger size and a longer lifetime.

\begin{figure*}
\vspace*{-80mm}
  \hspace*{-13mm}
 \includegraphics[width=200mm]{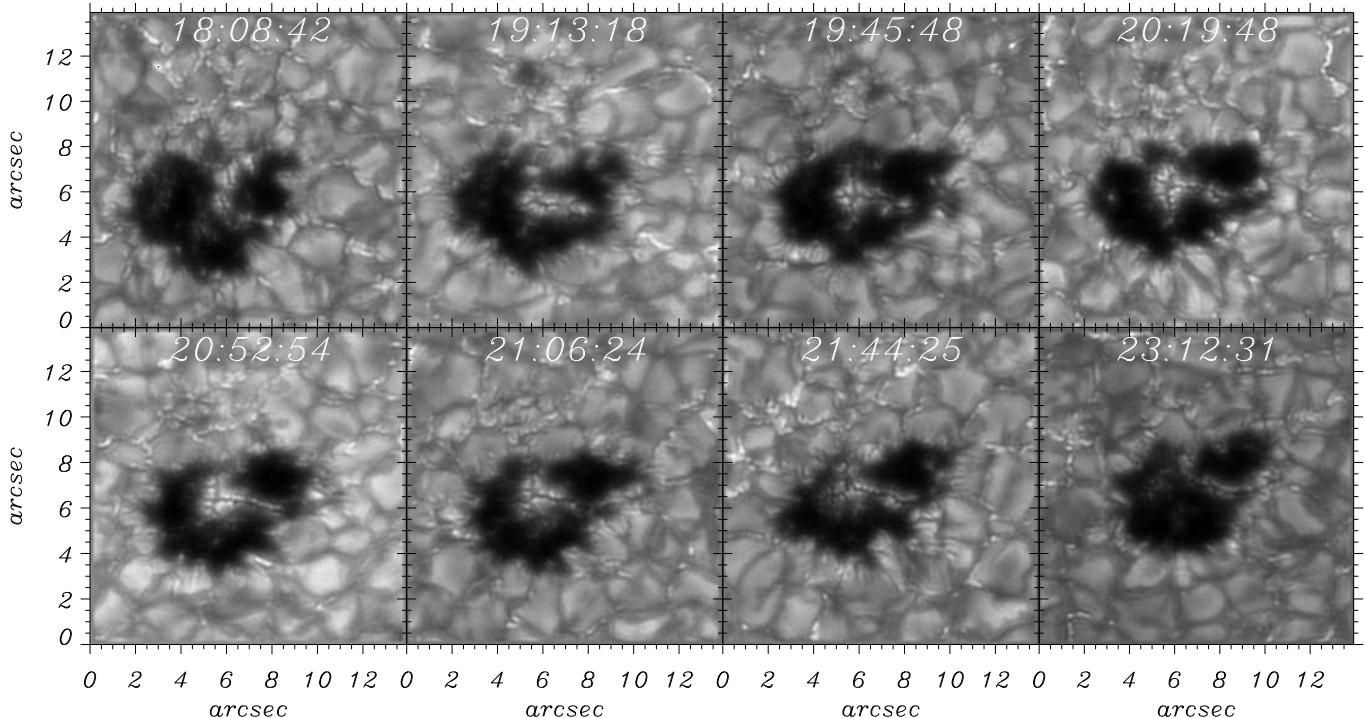}
 \vspace*{-57mm}
  \caption{Selected images of the pore from the G-band time series. The time increases from left to right and from top to bottom. Evolution of fine structure i.e. lightbridge fragments, lightbridge and larger cell can be recognised.}
\end{figure*}

\begin{figure*}
\vspace*{-30mm}
  \hspace*{-4mm}
 \includegraphics[width=180mm]{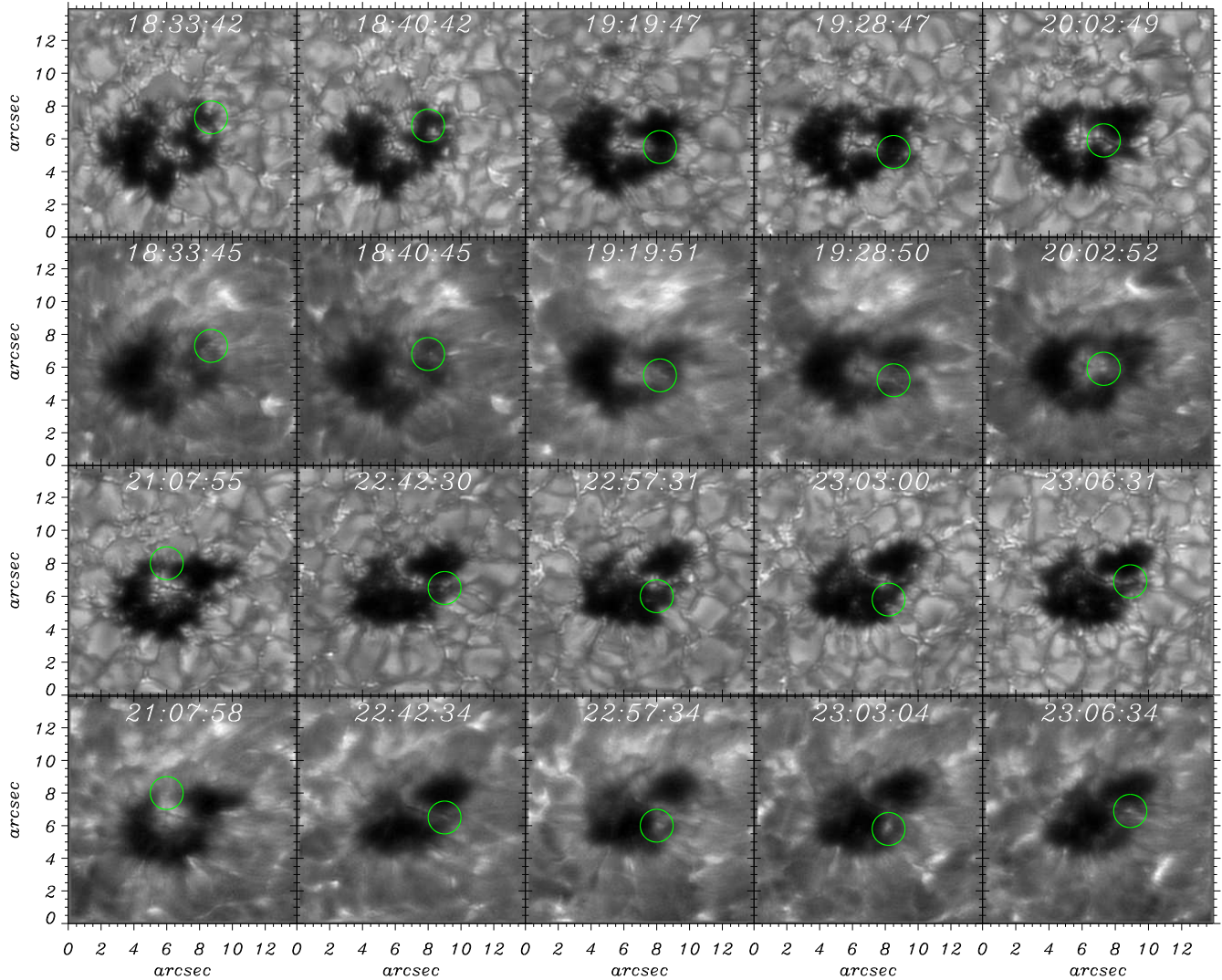}
 \vspace*{-12mm}
  \caption{Selected images that illustrates various short lived transient events in Ca II H. We highlight the location of the events of interest with a green circle. Co-spatial and nearly co-temporal G-band images are also displayed for photospheric reference. Time evolves from left to right and from top to bottom.}
\end{figure*}

\begin{figure*}
\vspace*{-95mm}
  \hspace*{-4mm}
 \includegraphics[width=180mm]{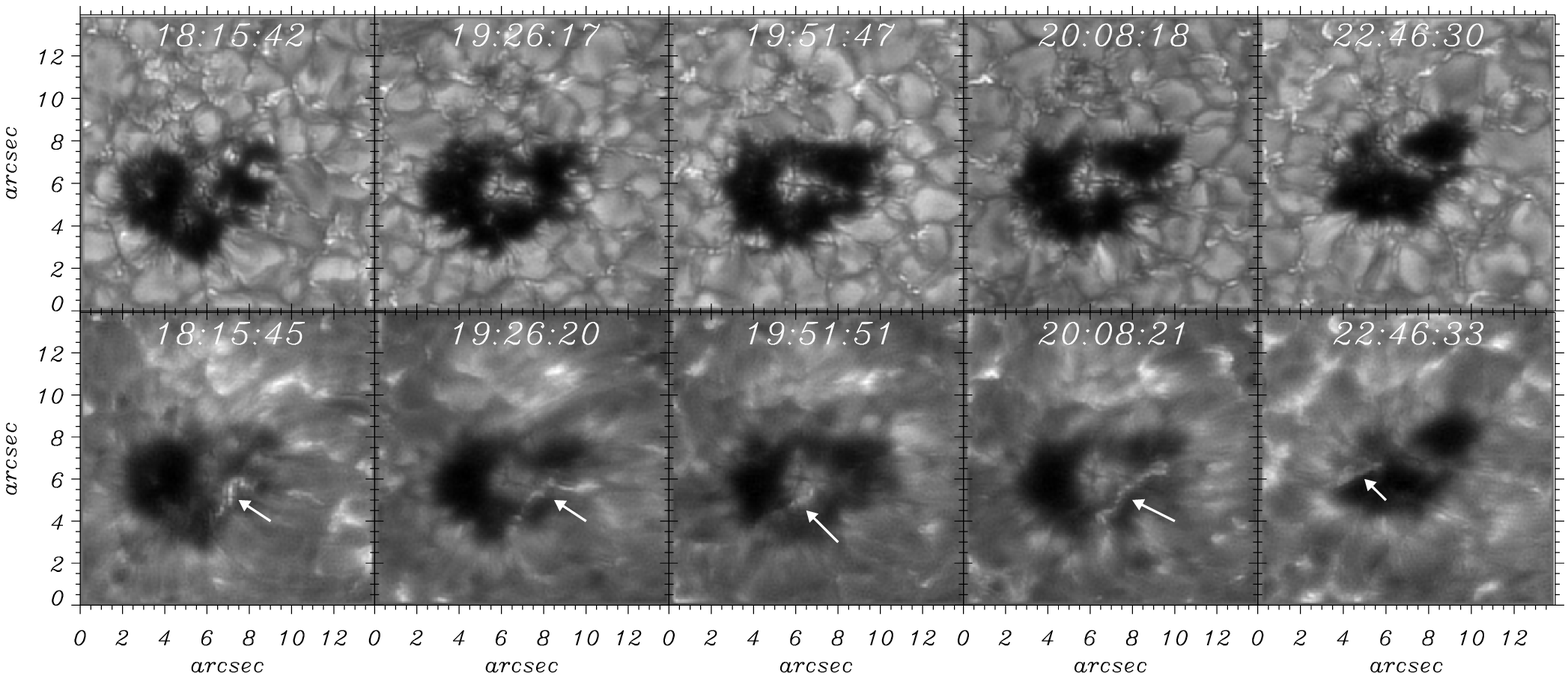}
 \vspace*{-12mm}
  \caption{Similar to Figure 3 but for ribbon like events. The location of the ribbon is indicated location by the white arrow.}
\end{figure*}

\subsection{Chromosphereic activity above pore}

The movie of the Ca II H time series reveals various transient and long lasting brightening events over the pore. All images are co-spatial
and nearly co-temporal to those taken on the G-band. We characterize such events in four categories according to their lifetime and structure. In the subsequent sections we discuss these events in detail.

\subsubsection{Short lived transient events}
From Ca II H time series we glaned 10 short lived transient events. Figure 3 highlights those events with co-spatial and co-temporal G-band and Ca II H broadband images. The events are marked with green circles in both wavebands. These events are dot like and elongated in shape. Most of the events (seven out of ten) are only observed in a single frame while there are three events that last longer and are detected on up to five consecutive frames. This implies that the cadence of the observations, i.e. 30 s, is not enough for tracking the evolution of these features. The underlying atmosphere, represented on the G-band images, suggests that some of the chromospheric events are located above or close to the photospheric UDs, the lightbridge and micro-filaments. Bharti et al. (2013) reported the presence of jet-like events in the sunspot umbra where the jet evolved from a dot-like structure and had a lifetime of around 1 min. Thus, these events seem to be different from those reported by Bharti et al. (2013) and \cite{nelson17} as seven events have a lifetime shorter than 1 minute. The events that last longer, with a lifetime from 1 to 2.5 min, are not jet-like features rather diffused dot-like. It is worth to mention that the width of the filter used for the Ca II H broadband observations is wide what allows a strong contribution from the photosphere. The fact that we can see the lightbridge and the large cells in Ca II H is a consequence of the mentioned filter width. This could be also the reason for the diffused appearance of last three events. On the other hand the large photospheric contribution present on the Ca II H images helps us to find the correspondence of these events with structures in the lower layers. The structural and lifetime difference from those reported from Bharti et al. (2013) may be because they are related to an upper photospheric phenomenon which is possible due to larger filter width. Thus, the seven events that are detected only in a single frame are different from the other three long-lasting events. A close inspection of the broadband images suggests that those events are real as any artifact e.g., the hit of a cosmic particle on the detector (CCD) would have different appearance in terms of multiple ring like structure in processed images and unusual larger brightness. However, we need data with better cadence and to improve the statistics of these features to confirm their existence as well as to provide a better explanation of their origin.

\subsubsection{Sheet or ribbon like events}

We also noticed five sheet or ribbon like events above the pore core structure. Four events were observed above the dark core and the bright large cell. Figure 4 illustrates those events with co-spatial and co-temporal G-band and Ca II H broadband images. Co-spatial and co-temporal G-band images are also displayed for photospheric reference. The white arrow on the Ca II H images points to the location of the mentioned events. The evolution of those events can be seen in the supplementary online material. They typically appear as a ribbon area whose brightness increases with time in the first phase and then decays monotonically after reaching its maximum brightness like flare ribbons seen in H$\alpha$. The life time of these sheet or ribbon ranges from about 3 to 12 min. Interestingly, we do not observe a noticeable ribbon body motion as it is observed during flares.
In the photosphere (see G-band images) there are umbral dots and a lightbridge in the vicinity of these ribbons.

\subsubsection{Brightenings around the pore boundary}

Figure 5 depicts the evolution of a ribbon, a jet and a brightening above the pore boundaries, a faint umbral dot and a micro penumbral filament. A ribbon starts to form around 23:05:34 UT (see the white arrow in the Ca II H panel). At 23:09:33 UT the formation of the ribbon is completed and it has achieved its maximum brightness. The location of the ribbon is around the edges of granules with dark striations towards the pore nucleus.  It starts to decay further and moves downwards left and a chain of bright dots formed around x=4\arcsec, y=7\arcsec as indicated by the pink arrow.
Such dots and associated jets were reported by \cite{robustini16} where the fan-shaped jets show bright dots along the light bridge jets. Generally, such bright structure in Ca II H correspond to photospheric bright points as seen in the G-band images at x=2\arcsec, y=9\arcsec. This confirms that the bright chain has a different origin than the G-band bright points and the filigree seen in the inter granular lanes. Similarly, a bright chain is also visible above the dark cored micro penumbral filament at 23:18:05 UT (indicated by the yellow arrow).

A bright blob above the faint umbral dot can be seen at 23:08:01 UT. The location of the bright blob is indicated by the green circle. The brightness of the blob increases with time it reaches its maximum brightness around 23:09:05 UT and then decreases. At 23:13:35 UT a jet like structure appears, directed downwards, seen from the blob (orange circle). The length and brightness of this jet increases with time and attains maximum brightness around 23:16:05 UT. Later its brightness decreases until it disappears at 23:19:34 UT.

We indicate with the cyan arrow the location of an elongated bright structure above a dark core penumbral microfilament at 23:19:34 UT. This brightning is seen only in a single image. A close inspection of the movie suggests that this brightning is not due to the migration of the bright penumbral grain in the photosphere but similar to the events discussed above, i.e. a different bright structure around the pore boundary that do not show such brightenings in the Ca II H images.

\begin{figure*}
\vspace*{-30mm}
  \hspace*{-4mm}
 \includegraphics[width=180mm]{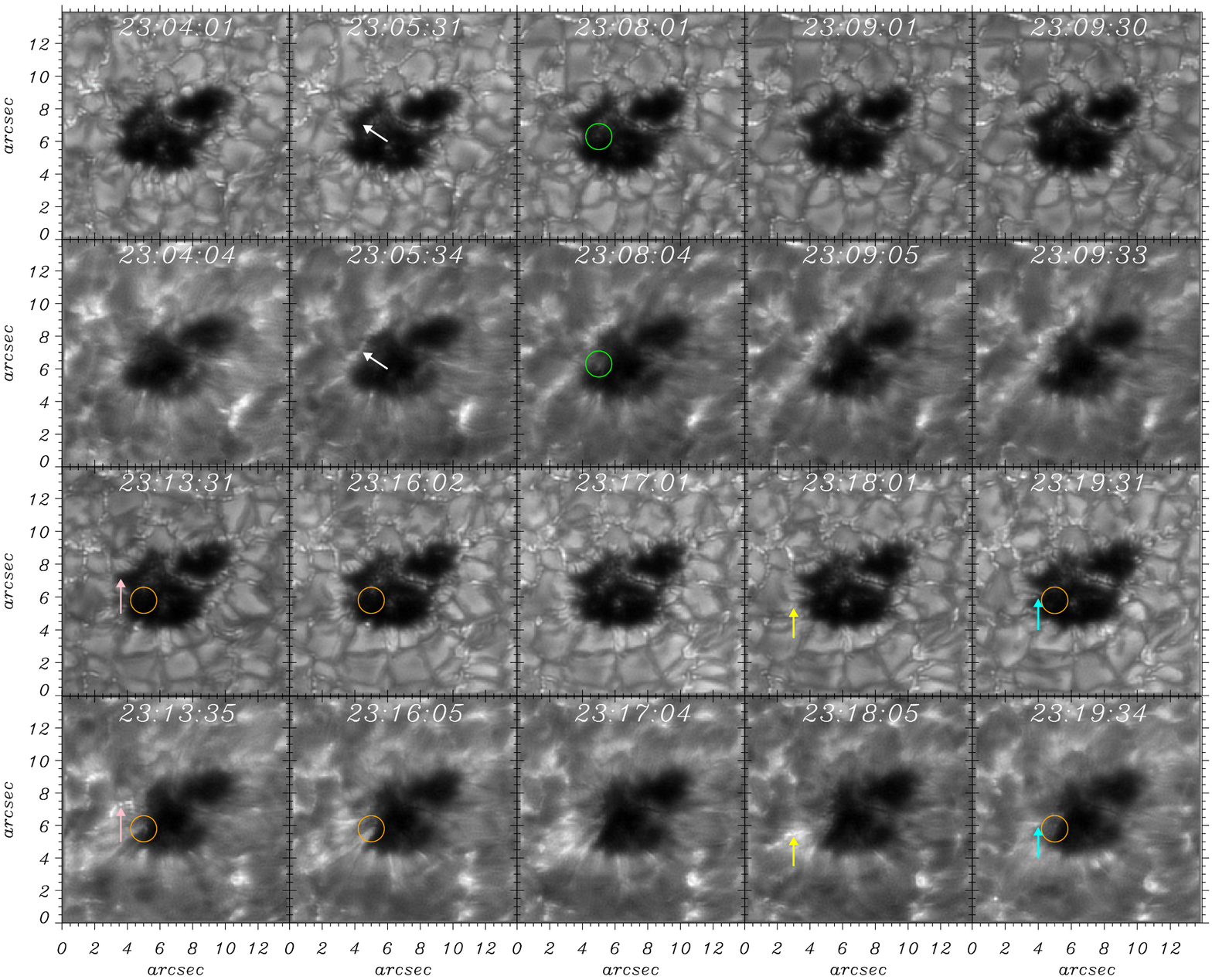}
 \vspace*{-12mm}
  \caption{Same as Figure 4 but for the brightenings around the pore boundary. Different events are highlighted with circles and arrows of different colors.}
\end{figure*}

\subsubsection{Jet like event above the larger cell}

Figure 6 demonstrates the plasma ejection event above the larger cell in the Ca II H and G-band
images in the photosphere. Before the event at 21:11 UT, a larger cell is visible in the photosphere. In the Ca II H images this larger cell is also visible. The clear visibility of the larger cell and its fine structure in both wavelength bands is due to the larger
filter width of Ca II H filter which has a strong contribution from the photosphere. This could explain why Bharti et al. (2013) did not observe any UD chromospheric signatures when analyzing Ca II images taken at the Swedish Solar Telescope (Scharmer et al. 2003) with a narrow filter width. At 21:13 UT a bright patch at x=5\arcsec and y=6\arcsec is visible in the Ca II H image. This patch evolves from x=7\arcsec and y=5\arcsec and moves towards the lower left part of the panel. This patch fades away and disappears at 21:15 UT. Around
21:16 UT another plasma ejection event starts above the larger UD that covers it. The evolution of this event and its photospheric counterpart are shown in the subsequent images of Figure 6. The area and intensity of the bright patches increases with time. At 21:20 UT we see three separate patches : two ribbon like patches located at $45^\circ$ apart and one roundish patch. Out of the two ribbon like patches one is short and narrow, while the other one is longer and wider. Both ribbons lived for about 11 minutes and crossed the pore boundaries. The roundish patch lived for about one minute and is visible only above the upper part of
the cell centered at x=7\farcs5 and y=5\farcs5.

We do not see any drastic changes in the large cell. However, we detect that the upper part of the cell at x=7\farcs5 and y=5\farcs5 starts to connect with the pore boundary through a lightbridge fragment that evolved at the edges of the dark structure. The typical motion found in the fine dark structures along the dark lane is visible in the photosphere.

\begin{figure*}
\vspace*{-30mm}
  \hspace*{-4mm}
 \includegraphics[width=180mm]{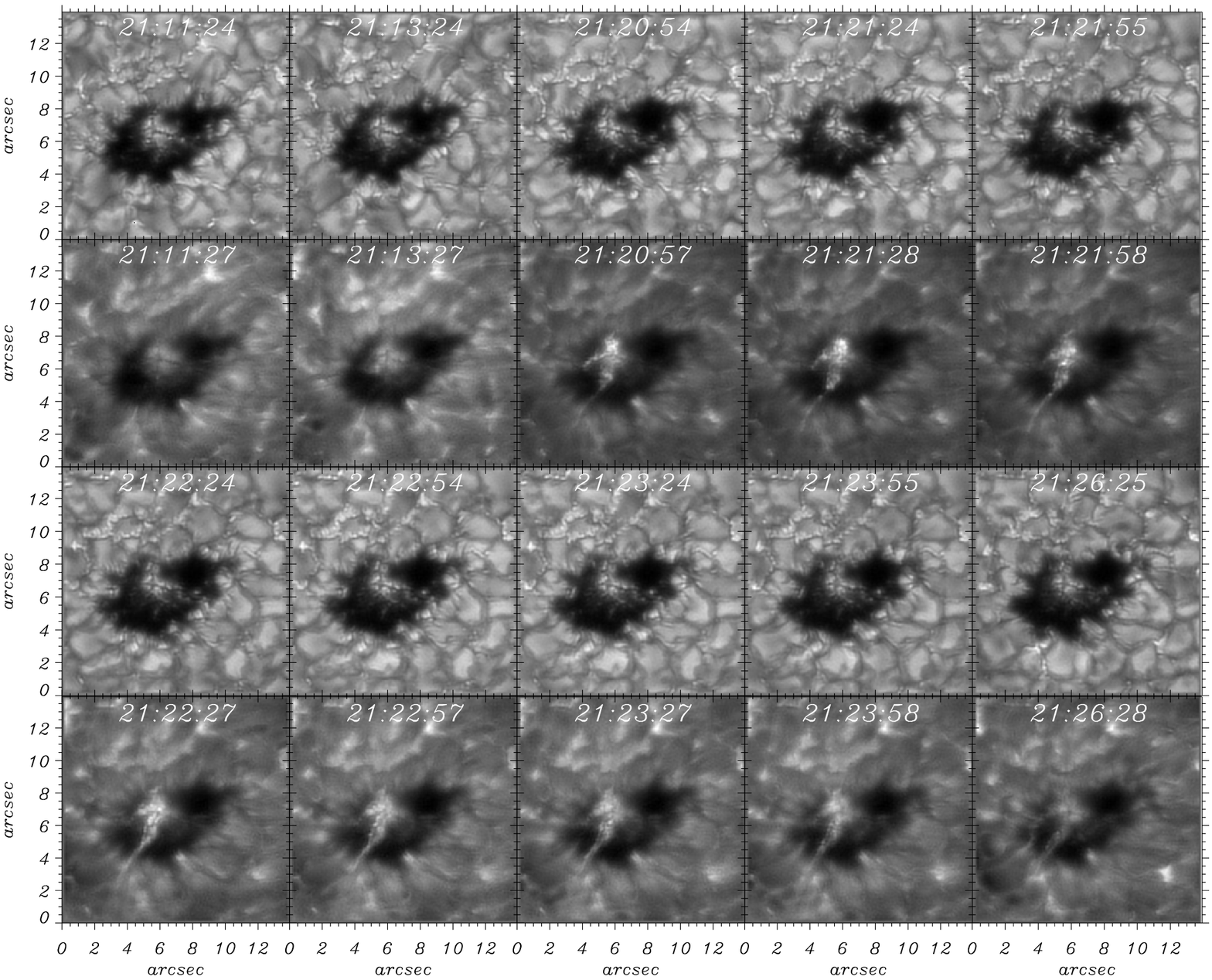}
 \vspace*{-12mm}
  \caption{Selected images that illustrates the evolution of the ejection event above the larger cell. First and third rows show G-band images.
  Second and fourth row display Ca II H images cospatial and nearly cotomporal to the G-band images. Time evolves from left to right and from top to bottom.}
\end{figure*}

\begin{figure*}
\vspace*{-600mm}
  \hspace*{-4mm}
 \centering
 \includegraphics[width=559mm]{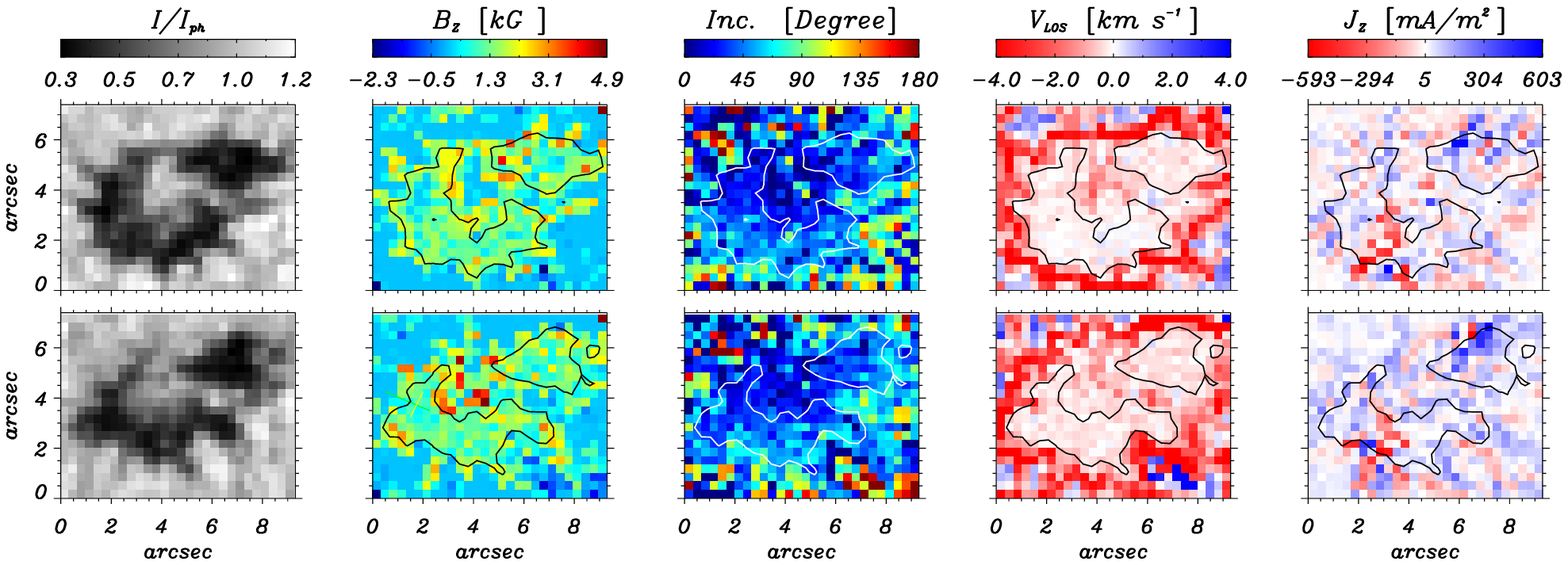}
 \vspace*{-9mm}
  \caption{Plasma parameters map at log$\tau$=0 derived from inversions before (top) and after (bottom) the ejection event above the larger cell. From left to right, we present the normalized continuum intensity, B$_{z}$, inclination, LOS velocity and vertical current density.}
\end{figure*}

\subsection{Plasma parameters before and after the ejection event}

Upper row of Figure 7 shows the plasma parameters derived  at log$\tau$=0 from the SP map before the ejection event that occurred between 21:02 and 21:04 UT. The data illustrates that the line of sight magnetic field in the large cell is lower (around 1500 G) in comparison with that of the dark part of the pore (more than 2000 G). This is in agreement with the findings of \cite{bharti09}. The magnetic diameter of the pore is higher than that found in continuum intensity observations \citep{keppens96,sobotka12}. The magnetic field is vertical in the center of the pore and becomes horizontal towards the edges of the pore. This is evident from the magnetic field inclination map. A ring of transverse field is visible around the pore where the filed is inclined up to $70^\circ$. The magnetic field
in the larger cell is slightly horizontal or inclined. These observed properties of the magnetic field vectors in the pore confirm findings of pervious works \citep{keppens96,sutter98,sankar03}. A similarly magnetic field vector in a larger cell is also in agreement with previous works \citep{bharti09}.
We do not see large differences between the LOS velocity on the bright and the dark parts of the structure. The darkest part of the pore was selected as velocity reference. However, downflow patches
are visible outside pore boundary \citep{hirzberger3,sobotka12}. We expect to detect upflowing material in the center part of the large UD when observing with a spatial resolution of 0\farcs64. However, we cannot observe them in this active region. We double checked the results using a line bisector analysis (not shown here) on the Fe I 6301.5 spectra line. The results of the depth dependent flows in the larger UD confirm that there is no significant difference between the large UD and the dark part of the pore. If the available spatial resolution is not enough to resolve the upflowing material then it can be mixed with different velocity contributions leading to different values or even opposite sign on the LOS velocity. The lower resolution of the SP data shows such larger downflow patches, which in fact are narrow downflow lanes. \cite{bharti19} demonstrated that at lower resolution such narrow downflow lanes shown up as large downflow patches.

Similar to \citep{shimizu09} and \citep{toriumi15b}, we also derived the vertical current density Jz from the inferred values of the horizontal components of the magnetic field vectors Bx and By (see, \cite{shimizu09}). Patches of the large positive and negative vertical current visible at the pore boundary, within the pore umbra and in the larger cell. According to
\cite{shimizu09} this could be the signature of a current sheet formation at the magnetic reconnection site. These are the locations from where the chromospheric plasma was ejected.

Similarly, the lower row in Figure 7 displays the plasma parameters derived from the SP map after the ejection event
over the larger cell from 21:44 to 21:46 UT at log$\tau$=0. The properties of the magnetic field vector are similar to those found before the ejection event.
No significant changes in the line of sight velocity are detected either. The only remarkable difference is observed in the Jz map where there are no patches of high current of any sign on the large cell. This could be due to a magnetic tension released after
the ejection event. It is evident from the Ca II H movie that no further ejection events are observed. On the other hand, there are patches of high current of both signs around x=2\arcsec and y=2\arcsec. This location corrosponds to that where we found the events describes in Section 3.2.3. The overall scenario for the magnetic field is similar to the one showed by penumbral filaments, i.e. the field strength isc weaker and more inclined respect to the LOS which makes a cusp above the overlying vertical magnetic field \citep{Rempel2011,tiwari13}.

\section{Discussions and conclusions}

We studied in detailed the chromospheric dynamics above a pore. To our knowledge this is the first time such event above the fine structure is studied and thoroughly described.  The G-band, Ca II H and SP observations followed
the evolution of the lightbridges and the larger cell as well as the spatial distribution of the magnetic field vector and LOS velocity within the pore. During 5 hours of observations the pore sustains its identity, i.e. it does not disintegrate or shows any signature of penumbra formation. The significant changes found within the pore were the formation of isolated bright features from lightbridge fragments that later transform into a larger cell. The migration of umbral dots in and away
from lightbridges have been reported by \cite{riethmuller08}. Realistic 3D MHD simulations of active region
formation by \cite{cheung10} show the formation of a lightbridge and umbral dots in a pore. \cite{Toriumi19} simulated a $\delta$-sunspot where a lightbridge forms between the umbra of opposite polarities. These simulations suggested that
penumbral filaments \citep{Rempel2011}, lightbridge and umbral dots have a similar nature and are produced by magnetoconvection. The migration of the dark structures from the center part of the larger cell towards the outer edges suggest mass motions of hot material that cools later, which is a typical signature of convection. \cite{bharti09} studied similar structures and found that they posses all the properties that umbral dots have.
Line of sight velocity maps do not show  such flow pattern in the larger structure. This could be an individual property of
 this structure. However, the inferred magnetic field strongly supports the hypothesis of magnetoconvection as the origin of this structure because it is weak and horizontal respect to the solar surface.

Realistic 3D MHD simulations presented in Schüssler \& Vögler (2006) were the first that proposed the convective nature of UDs. The simulations also produced jet-like outflows above the cusp in the upper atmosphere where upflow plumes are seen in the photosphere. However, due to the closed top boundary in those simulations, the details of the chromospheric dynamics remained elusive. These plumes or upflow jets could
be possibly caused by dynamic events (transient short lived events) in the chromospheric above sunspots or pore umbra. However, observational evidence are still elusive as reported in present study as well as the recent findings by \cite{bharti13} and \cite{nelson17}.

The important observational aspect is the the chromospheric plasma ejection above the larger UD.
If we assume that this structure is magnetoconvective in nature and compare with
simulations of \cite{bharti10b}, then the simulations predict strong downflows at the edges of the UDs. Those downflows can capture inclined magnetic field lines and drag them downwards, thereby creating a hairpin-like structure with patches of reversed polarity. Such opposite polarity system also occurs at the edges of penumbral filaments \citep{Rempel2011} and light bridges \citep{cheung10} . Thus, a close system of opposite polarity field forms eventually producing magnetic reconnetion what finally leads to the plasma ejection in the upper atmosphere. \cite{bharti07c} reported a surge event in a lightbridge of opposite polarity with umbra and suggested that such events
are caused by low altitude reconnection. Due to the lower spatial resolution of the SP maps used in this work
it is difficult to confirm such scenario. However jet like events observed in the umbra above umbral dots
reported by \cite{bharti13} support this mechanism.

\cite{shimizu09} and \cite{shimizu11} studied LB that produced chromospheric plasma ejection along its major axis. These authors were able to capture the information of the magnetic field and velocity vectors during the ejection event. They found downflows in the photosphere at the edge of the LB where the ejection event launched in the chromosphere. Moreover, high values of  opposite vertical current density were found at the ejection site in the photosphere. The authors suggested that this might be the detection of a flow in the current sheet that forms at the reconnection sites. Therefore, the observed patches of opposite vertical current density before the ejection in the large UD, as well as their disappearance after the ejection event, support a similar scenario of magnetic reconnection.

A similar mechanism has also been proposed to explain penumbral
microjets \citep[see][]{katsukawa07,magara10} where the reconnection take place between the vertical background field and the more inclined penumbral field.
Thus, the ejection might be caused by a reconnection between the pore's vertical field and the inclined one of the larger cell. According to \cite{toriumi15a}, the presence of patches of high vertical current of opposite  sign is indicative of magnetic shear which is a favorable condition for magnetic reconnection.

Jets above light bridges also generate small jet-like oscillations due to the p-mode leakage, which is sometimes called light walls \citep{yang15}. \cite{bharti15} reported that such oscillatory jets have bright leading edges. However, no such oscillatory behaviour was observed in the present study. \cite{keys18} demonstrated that magnetic pores can act as a waveguide that transmits significant wave energy to the upper atmosphere which can govern the dynamics of the lower solar atmosphere. The transmitted energy flux is significant to heat the localized atmosphere. The reported dynamics above the pore could have some relation with such oscillatory phenomenon and it is an open topic for further analysis. With better time resolution (1.4 s) Ca II H and Ca II  8542 full Stokes data \cite{nelson17} reported the presence of small-scale umbral brightenings similar to the  umbral-micro-jet event discovered by \cite{bharti13} and stated that such brightenings are not jet-like rather appear due to shock formation in the lower atmosphere. Ribbon like events over pore umbra are different from flares where two or more ribbons can be observed as a result of a magnetic reconnection in the upper atmosphere. The ribbon like events could be the result of low altitude reconnection caused by opposite polarity magnetic field at the edges of umbral dots which is not detected due to the limited resolution of telescope. Alternative, a ribbon could be the footpoint of coronal rain where heating may be caused by intense and impulsive chromospheric evaporation. It would be interesting to investigate if such a ribbon is also observed in H$\alpha$ above the umbra to confirm whether low altitude reconnection is the cause of coronal rain. It is well established that any small and large scale eruptions can be explained in terms of free magnetic energy contained in active regions which is accumulated in the form of electrical currents \citep{emslie12,asch14}. It has also been suggested that the magnetic field above sunspots umbra is complicated (Solanki 2003; Socas-Navarro 2005; Tritschler et al. 2008), thus different topologies and mechanisms could be possible for energy dissipation above umbra \citep{bharti13}. This scenario may also be valid for the umbra above pores.

 To confirm the reconnection scenario suggested by \cite{shimizu09}, \cite{shimizu11} and \cite{bharti10b} simultaneous
 observations in the photosphere and upper atmosphere with high resolution and high cadence are needed. This could be possible with state-of-the-art observations as those produced by the SST and that expected from DKIST (Keil et al. 2011) and EST (Collados et al.  2013)  in the near future. Data base of coordinated observations with SDO and IRIS will be useful.  On the other hand new inversion techniques proposed by \cite{vannoort12} and
  \cite{ruizcobo13}for SOT/SP can shed some light on such events.
\section*{Acknowledgments}

Hinode is a Japanese mission developed and launched by ISAS/JAXA,
with NAOJ as domestic partner and NASA and STFC (UK) as international partners.
It is operated by these agencies in co-operation with ESA and NSC (Norway).
This research is supported by Bal Shiksha Sadan Samiti (a nongovernmental organization [NGO]
, Udaipur). CQN was supported by the Research Council of Norway through its Centres of Excellence scheme, project number 262622.

\bibliographystyle{mn2e}
\bibliography{bharti}

\begin{thebibliography}{}

\bibitem[\protect\citeauthoryear{{Aschwanden}, {Xu} \& {Jing}}{{Aschwanden}
  et~al.}{2014}]{asch14}
{Aschwanden} M.~J.,  {Xu} Y.,    {Jing} J.,  2014, \apj, 797, 50

\bibitem[\protect\citeauthoryear{{Bharti}}{{Bharti}}{2015}]{bharti15}
{Bharti} L.,  2015, \mnras, 452, L16

\bibitem[\protect\citeauthoryear{{Bharti}, {Beeck} \& {Sch{\"u}ssler}}{{Bharti}
  et~al.}{2010}]{bharti10b}
{Bharti} L.,  {Beeck} B.,    {Sch{\"u}ssler} M.,  2010, \aap, 510, A12

\bibitem[\protect\citeauthoryear{{Bharti}, {Hirzberger} \& {Solanki}}{{Bharti}
  et~al.}{2013}]{bharti13}
{Bharti} L.,  {Hirzberger} J.,    {Solanki} S.~K.,  2013, \aap, 552, L1

\bibitem[\protect\citeauthoryear{{Bharti}, {Joshi} \& {Jaaffrey}}{{Bharti}
  et~al.}{2007}]{bharti07c}
{Bharti} L.,  {Joshi} C.,    {Jaaffrey} S.~N.~A.,  2007, \apjl, 669, L57

\bibitem[\protect\citeauthoryear{{Bharti}, {Joshi}, {Jaaffrey} \&
  {Jain}}{{Bharti} et~al.}{2009}]{bharti09}
{Bharti} L.,  {Joshi} C.,  {Jaaffrey} S.~N.~A.,    {Jain} R.,  2009, \mnras,
  393, 65

\bibitem[\protect\citeauthoryear{{Bharti}, {Quintero Noda}, {Joshi}, {Rakesh}
  \& {Pandya}}{{Bharti} et~al.}{2016}]{bharti16}
{Bharti} L.,  {Quintero Noda} C.,  {Joshi} C.,  {Rakesh} S.,    {Pandya} A.,
  2016, \mnras, 462, L93

\bibitem[\protect\citeauthoryear{{Bharti} \& {Rempel}}{{Bharti} \&
  {Rempel}}{2019}]{bharti19}
{Bharti} L.,  {Rempel} M.,  2019, arXiv e-prints, p. arXiv:1908.06439

\bibitem[\protect\citeauthoryear{{Bharti}, {Rimmele}, {Jain}, {Jaaffrey} \&
  {Smartt}}{{Bharti} et~al.}{2007}]{bharti07a}
{Bharti} L.,  {Rimmele} T.,  {Jain} R.,  {Jaaffrey} S.~N.~A.,    {Smartt}
  R.~N.,  2007, \mnras, 376, 1291

\bibitem[\protect\citeauthoryear{{Cheung}, {Rempel}, {Title} \&
  {Sch{\"u}ssler}}{{Cheung} et~al.}{2010}]{cheung10}
{Cheung} M.~C.~M.,  {Rempel} M.,  {Title} A.~M.,    {Sch{\"u}ssler} M.,  2010,
  \apj, 720, 233

\bibitem[\protect\citeauthoryear{{Dorotovi{\v{c}}}, {Rybansk{\'y}}, {Sobotka},
  {Lorenc}, {Barandas} \& {Fonseca}}{{Dorotovi{\v{c}}}
  et~al.}{2016}]{dorotovi16}
{Dorotovi{\v{c}}} I.,  {Rybansk{\'y}} M.,  {Sobotka} M.,  {Lorenc} M.,
  {Barandas} M.,    {Fonseca} J.~M.,  2016, in {Dorotovic} I.,  {Fischer}
  C.~E.,   {Temmer} M.,  eds, Coimbra Solar Physics Meeting: Ground-based Solar
  Observations in the Space Instrumentation Era Vol.~504 of Astronomical
  Society of the Pacific Conference Series, {Temporal Evolution of Magnetic
  Field and Intensity Properties of Photospheric Pores}.
p.~37

\bibitem[\protect\citeauthoryear{{Emslie}, {Dennis}, {Shih}, {Chamberlin},
  {Mewaldt}, {Moore}, {Share}, {Vourlidas} \& {Welsch}}{{Emslie}
  et~al.}{2012}]{emslie12}
{Emslie} A.~G.,  {Dennis} B.~R.,  {Shih} A.~Y.,  {Chamberlin} P.~C.,  {Mewaldt}
  R.~A.,  {Moore} C.~S.,  {Share} G.~H.,  {Vourlidas} A.,    {Welsch} B.~T.,
  2012, \apj, 759, 71

\bibitem[\protect\citeauthoryear{{Giordano}, {Berrilli}, {Del Moro} \&
  {Penza}}{{Giordano} et~al.}{2008}]{giordano08}
{Giordano} S.,  {Berrilli} F.,  {Del Moro} D.,    {Penza} V.,  2008, \aap, 489,
  747

\bibitem[\protect\citeauthoryear{{Hirzberger}}{{Hirzberger}}{2003}]{hirzberger3}
{Hirzberger} J.,  2003, \aap, 405, 331

\bibitem[\protect\citeauthoryear{{Ichimoto}, {Lites}, {Elmore}, {Suematsu},
  {Tsuneta}, {Katsukawa}, {Shimizu}, {Shine}, {Tarbell} \& {Title}}{{Ichimoto}
  et~al.}{2008}]{ichimoto08}
{Ichimoto} K.,  {Lites} B.,  {Elmore} D.,  {Suematsu} Y.,  {Tsuneta} S.,
  {Katsukawa} Y.,  {Shimizu} T.,  {Shine} R.,  {Tarbell} T.,    {Title} A.,
  2008, \solphys, 249, 233

\bibitem[\protect\citeauthoryear{{Katsukawa}, {Berger}, {Ichimoto}, {Lites},
  {Nagata}, {Shimizu}, {Shine}, {Suematsu}, {Tarbell}, {Title} \&
  {Tsuneta}}{{Katsukawa} et~al.}{2007}]{katsukawa07}
{Katsukawa} Y.,  {Berger} T.~E.,  {Ichimoto} K.,  {Lites} B.~W.,  {Nagata} S.,
  {Shimizu} T.,  {Shine} R.~A.,  {Suematsu} Y.,  {Tarbell} T.~D.,  {Title}
  A.~M.,    {Tsuneta} S.,  2007, Science, 318, 1594

\bibitem[\protect\citeauthoryear{{Keppens} \& {Martinez Pillet}}{{Keppens} \&
  {Martinez Pillet}}{1996}]{keppens96}
{Keppens} R.,  {Martinez Pillet} V.,  1996, \aap, 316, 229

\bibitem[\protect\citeauthoryear{{Keys}, {Morton}, {Jess}, {Verth}, {Grant},
  {Mathioudakis}, {Mackay}, {Doyle}, {Christian}, {Keenan} \&
  {Erd{\'e}lyi}}{{Keys} et~al.}{2018}]{keys18}
{Keys} P.~H.,  {Morton} R.~J.,  {Jess} D.~B.,  {Verth} G.,  {Grant} S. D.~T.,
  {Mathioudakis} M.,  {Mackay} D.~H.,  {Doyle} J.~G.,  {Christian} D.~J.,
  {Keenan} F.~P.,    {Erd{\'e}lyi} R.,  2018, \apj, 857, 28

\bibitem[\protect\citeauthoryear{{Lites}, {Akin}, {Card}, {Cruz}, {Duncan},
  {Edwards}, {Elmore}, {Hoffmann}, {Katsukawa}, {Katz}, {Kubo}, {Ichimoto},
  {Shimizu}, {Shine}, {Streander}, {Suematsu}, {Tarbell}, {Title} \&
  {Tsuneta}}{{Lites} et~al.}{2013}]{splites13}
{Lites} B.~W.,  {Akin} D.~L.,  {Card} G.,  {Cruz} T.,  {Duncan} D.~W.,
  {Edwards} C.~G.,  {Elmore} D.~F.,  {Hoffmann} C.,  {Katsukawa} Y.,  {Katz}
  N.,  {Kubo} M.,  {Ichimoto} K.,  {Shimizu} T.,  {Shine} R.~A.,  {Streander}
  K.~V.,  {Suematsu} A.,  {Tarbell} T.~D.,  {Title} A.~M.,    {Tsuneta} S.,
  2013, \solphys, 283, 579

\bibitem[\protect\citeauthoryear{{Magara}}{{Magara}}{2010}]{magara10}
{Magara} T.,  2010, \apjl, 715, L40

\bibitem[\protect\citeauthoryear{{Nelson}, {Henriques}, {Mathioudakis} \&
  {Keenan}}{{Nelson} et~al.}{2017}]{nelson17}
{Nelson} C.~J.,  {Henriques} V.~M.~J.,  {Mathioudakis} M.,    {Keenan} F.~P.,
  2017, \aap, 605, A14

\bibitem[\protect\citeauthoryear{{Ortiz}, {Bellot Rubio} \& {Rouppe van der
  Voort}}{{Ortiz} et~al.}{2010}]{ortiz10}
{Ortiz} A.,  {Bellot Rubio} L.~R.,    {Rouppe van der Voort} L.,  2010, \apj,
  713, 1282

\bibitem[\protect\citeauthoryear{{Quintero Noda}, {Asensio Ramos}, {Orozco
  Su{\'a}rez} \& {Ruiz Cobo}}{{Quintero Noda} et~al.}{2015}]{qnoda15}
{Quintero Noda} C.,  {Asensio Ramos} A.,  {Orozco Su{\'a}rez} D.,    {Ruiz
  Cobo} B.,  2015, \aap, 579, A3

\bibitem[\protect\citeauthoryear{{Quintero Noda}, {Shimizu} \&
  {Suematsu}}{{Quintero Noda} et~al.}{2016}]{qnoda16a}
{Quintero Noda} C.,  {Shimizu} T.,    {Suematsu} Y.,  2016, \mnras, 457, 1703

\bibitem[\protect\citeauthoryear{{Rempel}}{{Rempel}}{2011}]{Rempel2011}
{Rempel} M.,  2011, \apj, 729, 5

\bibitem[\protect\citeauthoryear{{Riethm{\"u}ller}, {Solanki}, {Zakharov} \&
  {Gandorfer}}{{Riethm{\"u}ller} et~al.}{2008}]{riethmuller08}
{Riethm{\"u}ller} T.~L.,  {Solanki} S.~K.,  {Zakharov} V.,    {Gandorfer} A.,
  2008, \aap, 492, 233

\bibitem[\protect\citeauthoryear{{Rimmele}}{{Rimmele}}{2008}]{rimmele08}
{Rimmele} T.,  2008, \apj, 672, 684

\bibitem[\protect\citeauthoryear{{Robustini}, {Leenaarts}, {de la Cruz
  Rodriguez} \& {Rouppe van der Voort}}{{Robustini} et~al.}{2016}]{robustini16}
{Robustini} C.,  {Leenaarts} J.,  {de la Cruz Rodriguez} J.,    {Rouppe van der
  Voort} L.,  2016, \aap, 590, A57

\bibitem[\protect\citeauthoryear{{Ruiz Cobo} \& {Asensio Ramos}}{{Ruiz Cobo} \&
  {Asensio Ramos}}{2013}]{ruizcobo13}
{Ruiz Cobo} B.,  {Asensio Ramos} A.,  2013, \aap, 549, L4

\bibitem[\protect\citeauthoryear{{Sankarasubramanian} \&
  {Rimmele}}{{Sankarasubramanian} \& {Rimmele}}{2003}]{sankar03}
{Sankarasubramanian} K.,  {Rimmele} T.,  2003, \apj, 598, 689

\bibitem[\protect\citeauthoryear{{Shimizu}}{{Shimizu}}{2011}]{shimizu11}
{Shimizu} T.,  2011, \apj, 738, 83

\bibitem[\protect\citeauthoryear{{Shimizu}, {Katsukawa}, {Kubo}, {Lites},
  {Ichimoto}, {Suematsu}, {Tsuneta}, {Nagata}, {Shine} \& {Tarbell}}{{Shimizu}
  et~al.}{2009}]{shimizu09}
{Shimizu} T.,  {Katsukawa} Y.,  {Kubo} M.,  {Lites} B.~W.,  {Ichimoto} K.,
  {Suematsu} Y.,  {Tsuneta} S.,  {Nagata} S.,  {Shine} R.~A.,    {Tarbell}
  T.~D.,  2009, \apjl, 696, L66

\bibitem[\protect\citeauthoryear{{Sobotka}, {Bonet} \& {Vazquez}}{{Sobotka}
  et~al.}{1993}]{sobotka93}
{Sobotka} M.,  {Bonet} J.~A.,    {Vazquez} M.,  1993, \apj, 415, 832

\bibitem[\protect\citeauthoryear{{Sobotka}, {Bonet} \& {Vazquez}}{{Sobotka}
  et~al.}{1994}]{sobotka94}
{Sobotka} M.,  {Bonet} J.~A.,    {Vazquez} M.,  1994, \apj, 426, 404

\bibitem[\protect\citeauthoryear{{Sobotka}, {Del Moro}, {Jur{\v c}{\'a}k} \&
  {Berrilli}}{{Sobotka} et~al.}{2012}]{sobotka12}
{Sobotka} M.,  {Del Moro} D.,  {Jur{\v c}{\'a}k} J.,    {Berrilli} F.,  2012,
  \aap, 537, A85

\bibitem[\protect\citeauthoryear{{Sobotka}, {{\v{S}}vanda}, {Jur{\v{c}}{\'a}k},
  {Heinzel}, {Del Moro} \& {Berrilli}}{{Sobotka} et~al.}{2013}]{sobotka13}
{Sobotka} M.,  {{\v{S}}vanda} M.,  {Jur{\v{c}}{\'a}k} J.,  {Heinzel} P.,  {Del
  Moro} D.,    {Berrilli} F.,  2013, \aap, 560, A84

\bibitem[\protect\citeauthoryear{{Suetterlin}}{{Suetterlin}}{1998}]{sutter98}
{Suetterlin} P.,  1998, \aap, 333, 305

\bibitem[\protect\citeauthoryear{{Tiwari}, {van Noort}, {Lagg} \&
  {Solanki}}{{Tiwari} et~al.}{2013}]{tiwari13}
{Tiwari} S.~K.,  {van Noort} M.,  {Lagg} A.,    {Solanki} S.~K.,  2013, \aap,
  557, A25

\bibitem[\protect\citeauthoryear{{Toriumi}, {Cheung} \& {Katsukawa}}{{Toriumi}
  et~al.}{2015}]{toriumi15b}
{Toriumi} S.,  {Cheung} M. C.~M.,    {Katsukawa} Y.,  2015, \apj, 811, 138

\bibitem[\protect\citeauthoryear{{Toriumi} \& {Hotta}}{{Toriumi} \&
  {Hotta}}{2019}]{Toriumi19}
{Toriumi} S.,  {Hotta} H.,  2019, \apjl, 886, L21

\bibitem[\protect\citeauthoryear{{Toriumi}, {Katsukawa} \& {Cheung}}{{Toriumi}
  et~al.}{2015}]{toriumi15a}
{Toriumi} S.,  {Katsukawa} Y.,    {Cheung} M. C.~M.,  2015, \apj, 811, 137

\bibitem[\protect\citeauthoryear{{Tsuneta}, {Ichimoto}, {Katsukawa}, {Nagata},
  {Otsubo}, {Shimizu}, {Suematsu}, {Nakagiri}, {Noguchi}, {Tarbell}, {Title},
  {Shine} \& {Rosenberg}}{{Tsuneta} et~al.}{2008}]{tsuneta08}
{Tsuneta} S.,  {Ichimoto} K.,  {Katsukawa} Y.,  {Nagata} S.,  {Otsubo} M.,
  {Shimizu} T.,  {Suematsu} Y.,  {Nakagiri} M.,  {Noguchi} M.,  {Tarbell} T.,
  {Title} A.,  {Shine} R.,    {Rosenberg} W.,  2008, \solphys, 249, 167

\bibitem[\protect\citeauthoryear{{van Noort}}{{van Noort}}{2012}]{vannoort12}
{van Noort} M.,  2012, \aap, 548, A5

\bibitem[\protect\citeauthoryear{{Verma} \& {Denker}}{{Verma} \&
  {Denker}}{2014}]{verma14}
{Verma} M.,  {Denker} C.,  2014, \aap, 563, A112

\bibitem[\protect\citeauthoryear{{Yang}, {Zhang}, {Jiang} \& {Xiang}}{{Yang}
  et~al.}{2015}]{yang15}
{Yang} S.,  {Zhang} J.,  {Jiang} F.,    {Xiang} Y.,  2015, \apjl, 804, L27

\end{thebibliography}
\end{document}